\documentclass[a4paper,11pt]{article}
\pdfoutput=1 

\usepackage{jcappub} 

\usepackage[T1]{fontenc} 

\title{\boldmath Excess of lensing amplitude in the Planck CMB power spectrum}

\author[1]{Rahima Mokeddem,\note{Corresponding author.}}
\author[2]{Wiliam S. Hipólito-Ricaldi,}
\author[3]{Armando Bernui}

\affiliation[1]{PPGCosmo, CCE, Universidade Federal do Esp\'{\i}rito Santo (UFES), Av. Fernando Ferrari, 540, CEP 29.075-910, Vitória, ES, Brazil.}
\affiliation[2]{N\'ucleo Cosmo-UFES \& Departamento de Ci\^encias Naturais, Universidade Federal do Esp\'{\i}rito Santo (UFES), 	 Rodovia BR 101 Norte,  km. 60, CEP 29.932-540, S\~ao Mateus, ES, Brazil.}
\affiliation[3]{Observat\'orio Nacional, Rua General Jos\'e Cristino 77, 
	S\~ao Crist\'ov\~ao, CEP 20921-400 Rio de Janeiro, RJ, Brazil.}

\emailAdd{rahima.mokeddem@edu.ufes.br}

\abstract{Precise measurements of the Planck cosmic microwave background (CMB) angular power spectrum (APS) at small angles have stimulated 
accurate statistical analyses of the lensing amplitude parameter $A_{L}$. To confirm if it satisfies the value expected by the flat-$\Lambda$CDM concordance model, i.e. $A_{L} = 1$, 
we investigate the spectrum difference obtained as the difference of the measured Planck CMB APS and the Planck best-fit 
$\Lambda$CDM APS model. To know if this residual spectrum corresponds to statistical noise or if it has a hidden signature that can be accounted for 
with a larger lensing amplitude $A_{L} > 1$, 
we apply the Ljung-Box statistical test and find, with high statistical significance, that the spectrum difference is not statistical noise. 
This spectrum difference is then analysed in detail using simulated APS, based on the Planck $\Lambda$CDM best-fit model, where the 
lensing amplitude is a free parameter. 
We explore different binnations of the multipole order \,$\ell$\, and look for the best-fit lensing amplitude parameter that accounts 
for the spectrum difference in a $\chi^2$ procedure. 
We find that there is an excess of signal that is well explained by a $\Lambda$CDM APS with a non-null lensing amplitude parameter $A_{lens}$, with values in the interval $[0.10,0.29]$ at 68\% confidence level. 
Furthermore, the lensing parameter in the Planck APS should be $1 + A_{lens} > 1$ at $\sim 3 \sigma$ of  statistical confidence. 
Additionally, we perform statistical tests that confirm the robustness of this result. 
Important to say that this excess of lensing amplitude, not accounted in the Planck's flat-$\Lambda$CDM model, could have an impact 
on the theoretical expectation of large-scale structures formation once the scales where it was detected correspond to these matter 
clustering processes.}

\begin{document}
\maketitle
\flushbottom

\section{Introduction}
\label{sec:intro}

The current standard model of cosmology was recently confirmed analysing the precise cosmic microwave background (CMB) 
radiation data measured by the Planck satellite~\citep{Planck18}. 
The 6-parameters cosmological model that best fits the observational data, from CMB radiation and other cosmological observables, is the flat-$\Lambda$CDM, where the 3-dimensional space is Euclidean (i.e., zero spatial curvature or $k = 0$), and the universe is filled in with cold dark matter (CDM) and dark energy –in the form of a cosmological constant $\Lambda$– in addition to the standard baryonic and electromagnetic ingredients~\citep{Planck18,Planck16}.

One distinctive feature of the Planck measurements is the high angular resolution that has allowed to measure the CMB temperature fluctuations at the scales $\sim 4'$ (i.e., $\ell \sim 2500$). 
This made it possible to reconstruct, by inverse procedure, the lensing potential map that causes the weak lensing action of matter on the CMB photons path from the last scattering surface to us~\citep{Planck15-XV,Jia15,GAM,Bianchini}. 
In fact, the CMB photons detected by Planck are already weakly lensed by the matter they encounter in their paths, and the prediction of this phenomenon in the standard model is a tiny effect of smoothing the CMB acoustic  peaks at very small angular scales (i.e., $\ell \stackrel{>}{_{\sim}} 1000$). 
Since the observed CMB photons can not be delensed, the solution adopted by the Planck collaboration for the CMB data analyses is ma the weak lensing effect in the model that best fits the CMB angular power spectrum (APS). In practice, this solution contemplates that the amplitude of the weak lensing effect on CMB photons is, in the 
flat-$\Lambda$CDM model, $A_{L} = 1$.

The precise measurements of the Planck  CMB temperature-temperature (TT) APS at small angles have stimulated 
accurate statistical analyses about the suitability of the lensing amplitude parameter $A_{L} = 1$, 
as a way to validate the $\Lambda$CDM concordance model. 
As a result, some works have reported discordance with this amplitude. 
In fact, previous analyses indicate a clear preference for a higher lensing amplitude, i.e.,  $A_{L} > 1$
\citep{Calabrese,Bianchini,DES22,Ballardini,Valentino20b,Valentino20a,Planck16}\footnote{see the table 9 in page 32, in  ref. 
https://arxiv.org/pdf/1605.02985.pdf}. 
More recently, the Planck  collaboration reported: 
$A_{L} = 1.180 \,\pm\, 0.065$ (68\%, Planck TT,TE,EE+lowE) and 
$A_{L} = 1.243 \,\pm\, 0.096$ (68 \%, Planck TT+lowE), where the lensing amplitude $A_L$ is varied in the 
parameter analysis (see section 6.2 in~\cite{Planck18}). 
These results motivate us to perform a detailed 
examination of the CMB acoustic peaks sensitive to the lens phenomenon, i.e. $\ell > 1000$, 
due to a possible excess of lensing power in the TT APS data, a signal that can be accounted for with a 
lensing amplitude $A_{L} > 1$, that is, higher than the value adopted by the flat-$\Lambda$CDM Planck's 
best-fit model $A_{L} = 1$. 

The outline of this work is the following: 
In section~\ref{sec2} we describe the Planck data employed in this work. 
In section~\ref{sec3} we perform the statistical analyses of the spectra difference (Planck CMB APS minus Planck best-fit 
$\Lambda$CDM APS) to investigate: 
(i) if it corresponds to white noise or not; 
in negative case, (ii)  if it has signature and amplitude that can be explained by an extra lensing amplitude $A_{lens} > 0$ 
interpreted as a possible excess of lensing signature not accounted by the Planck  best-fit $\Lambda$CDM APS; 
(iii) if this spectrum difference can be explained by modifying some other cosmological parameters in the APS of 
the $\Lambda$CDM model. 
In section~\ref{sec4} we discuss our results, and also present our conclusions and final remarks.

\section{Data}\label{sec2}

The most accurate current measurements of the CMB temperature fluctuations are part of the third public data release 
of the Planck collaboration~\cite{Planck18}, precise data that allow to re-examine many interesting features already reported 
with diverse CMB data  sets~\citep{WHR1,B-HR,Samal,BR12,Aluri12,BOP,Polastri,Aluri16,Aluri17,Vafaei,Goyal,Saha21,Chiocchetta,Saha22,Owusu}. 
Further, CMB products combined with data sets from other cosmological tracers are being used to study models alternative to the flat-$\Lambda$CDM model (see, e.g.~\cite{WHR2,WHR3,Bessa,Nunes}).

In particular, the large set of data products released by the Planck team offers the possibility to perform exhaustive analyses 
to learn more about features of the matter clustering~\citep{Marques20a,Avila1,Avila2}, investigating the small angular scales where the weak lensing 
phenomenon left its signature imprinted. In fact, the Planck CMB data are specially accurate at small angular scales due to their high angular 
resolution~\citep{Marques20b,Dong,Tanseri}. 
Thus, we shall analyse in details the small angular scales of the CMB TT APS which is part 
of the third data release of the PLANCK collaboration~\cite{Planck18}.  Our main goal is to study if the difference between the measured Planck APS and the best-fit $\Lambda$CDM APS (based on the 6-parameters flat-$\Lambda$CDM model) 
from the data fit analyses done by the Planck team, is just statistical noise or it has indeed some indicative signature of an incorrect modeling of the weak lensing amplitude (i.e., $A_L \ne 1$), or of other cosmological parameters. 
Clearly, the null hypothesis that deserves investigation establishes  that the difference between the observed Planck APS 
and the $\Lambda$CDM APS is just white noise.

\vspace{0.3cm}
For our analyses some features are important to be considered. 
Firstly, the possible signal would be significantly detected only at the angular scales with the smallest 
error measurements, that is, for the interval $\ell \sim 1000 - 2000$; 
at the same time, one knows that these scales correspond to the regime where the lensing phenomenon has a large impact on 
the CMB APS. 
Secondly, for these analyses we use the original unbinned data from Planck repository pages, then our analyses shall consider 
binnation inspections beside  the  $\Delta \ell = 30$ case  considered by the Planck team. 

In our scrutiny, the released Planck CMB TT APS, the Planck APS~\footnote{COM$_{-}$PowerSpect$_{-}$CMB-TT-full$_{-}$R3.01.txt}, 
covers the range of multipoles $\ell = 2-2508$, where the multipoles of interest here, $\ell \ge 30$, were derived from the 
cross-half-mission likelihood Planck, Plik (for details see ref.~\cite{Planck18-V}). 
The Planck APS was derived from the Commander component-separation algorithm applied to the combination of Planck 2018 temperature 
data between 30 and 857 GHz, including 86\% of the sky (Planck-2018 results VI) (associated $1\sigma$ errors include beam uncertainties).

The analysis was done by optimally combining the spectra in the frequency range $100-217$ GHz, and correcting them for unresolved 
foregrounds using the best-fit foreground solution (regions were chosen as to avoid the areas dominated by noise). 
For the best-fit $\Lambda$CDM CMB APS, to be subtracted to the Planck APS for the aim of our analyses, we use the 6-parameter 
flat-$\Lambda$CDM minimal model, also released by the Planck team and hereafter termed the $\Lambda$CDM 
APS~\footnote{COM$_{-}$PowerSpect$_{-}$CMB-base-plikHM-TTTEEE-lowl-lowE-lensing-minimum-theory$_{-}$R3.01.txt}. 
This $\Lambda$CDM APS uses the Planck TT, TE, EE+lowE+lensing data. 
Both power spectra, the Planck APS and $\Lambda$CDM APS, were publicly released by the Planck collaboration\footnote{
\url{https://archives.esac.esa.int/doi/html/data/astronomy/planck/Cosmology.html}
}.

\section{Methodology and data analyses}\label{sec3}

This section describes the methodological approach we shall follow to study the Planck data APS, specifically concerning the 
possibility that the lensing amplitude parameter that best-fits the APS data is larger than 1, $A_L > 1$, where the value 
used by the Planck collaboration to  obtain the flat-$\Lambda$CDM model is $A_L = 1$.

\subsection{Methodology to investigate the spectrum difference}\label{sec3.1}

We want to investigate if there is an excess of lensing power in the TT APS measured by the Planck Collaboration 
$C^{\mbox{\footnotesize Planck}}_{\ell}$, with 1$\sigma$ standard deviation $\sigma^{\mbox{\footnotesize Planck}}_{\ell}$, 
with respect to the flat-$\Lambda$CDM APS, $C^{\Lambda\mbox{\footnotesize CDM}}_{\ell}$, 
obtained through a best-fit procedure of the 6 parameter flat-$\Lambda$CDM cosmological model to the TT Planck APS data~\citep{Planck18}. 
For this, we first define the spectrum difference :
\begin{equation}\label{Dobs}
\delta^{obs}_{\ell} \equiv \frac{C^{\mbox{\footnotesize Planck}}_{\ell} - C^{\Lambda\mbox{\footnotesize CDM}}_{\ell}} 
{C^{\Lambda\mbox{\footnotesize CDM}}_{\ell}} \,. 
\end{equation}
Notice that the $\Lambda$CDM APS,  $C^{\Lambda\mbox{\footnotesize CDM}}_{\ell}$, takes into account the CMB lensing phenomenon with $A_L = 1$, which is the value assumed by the Planck Collaboration in the  flat-$\Lambda$CDM model.

We define the symbol $A_{lens} \in [0,1]$
as the parameter that quantifies the excess of lensing amplitude to explain the residual $\delta^{obs}_{\ell}$. 
We construct synthetic APS,  $C^{syn,L}_{\ell}(A_{lens})$, 
with the Planck 2018 flat-$\Lambda$CDM cosmological parameters but for arbitrary lensing amplitude 
$A_{lens} \ge 0$. 
We also define
\begin{equation}\label{Dsyn} 
\delta^{\mbox{\,\footnotesize exc}}_{\ell}(A_{lens}) \,\equiv\, \frac{ C^{\mbox{\,\footnotesize syn},L}_{\ell}(A_{lens}) 
- C^{\mbox{\,\footnotesize syn},uL}_{\ell} } {C^{\mbox{\,\footnotesize syn},uL}_{\ell}} \,,
\end{equation}
where 
\begin{eqnarray}
C^{\Lambda\mbox{\footnotesize CDM}}_{\ell} &=& C^{\mbox{\,\footnotesize syn},L}_{\ell}(A_{lens} = 1) \, , \\
C^{\mbox{\,\footnotesize syn},uL}_{\ell} &=& C^{\mbox{\,\footnotesize syn},L}_{\ell}(\mbox{A}_{lens}=0) \,,
\end{eqnarray}
where the upper letters $L$ and $uL$ mean {\em lensed} and {\em unlensed} APS, respectively.

%
%
%
The quantity $\delta^{\mbox{\,\footnotesize exc}}_{\ell}(A_{lens})$  measures the possible excess of lensing power 
represented by the relative difference of synthetic lensed and unlensed APS (the difference that depends on the parameter 
$A_{lens}$; clearly, the case $A_{lens} = 0$ implies 
that $\delta^{\mbox{\,\footnotesize exc}}_{\ell} = 0$). 
Then, a good statistical agreement between the excess quantity $\delta^{\mbox{\,\footnotesize exc}}_{\ell}$ and the observed difference 
$\delta^{\mbox{\,\footnotesize obs}}_{\ell}$, measured with
the reduced $\chi^2$ best-fit for some $A_{lens} > 0$ value, 
will provide an explanation for $\delta^{\mbox{\,\footnotesize obs}}_{\ell}$ as being an excess of lensing power in the Planck APS not accounted for
by the $\Lambda$CDM lensing amplitude $A_{L} = 1$. 
Of course, other interpretations for $\delta^{\mbox{\,\footnotesize obs}}_{\ell}$ would be possible, as for instance that it is just statistical noise, and for this reason it is interesting to examine them too. In order to evaluate if the spectrum difference $\delta^{\mbox{\,\footnotesize obs}}_{\ell}$ corresponds to white noise,  we shall apply the Ljung-Box test \citep{ljungbox}. 

Our analyses include the simulation of APS, from now on termed synthetic APS, which considers the modeling of 
the weak lensing phenomenon on CMB photons for cases with $A_{lens} = 0$ (unlensed) and $A_{lens} \ne 0$ (lensed). 
The details for the production of these synthetic APS are the following: 

\begin{itemize}
\item The Boltzmann code  $CLASS$  \citep{class} was modified to allow the code to consider 
the lensing amplitude $A_{lens}$ as a parameter. 

\item
We use this modified code with the cosmological parameters corresponding to the observed Planck TT APS, as detailed in the 
table~\ref{table1}, to produce synthetic lensed and unlensed $\Lambda$CDM TT APS, $C^{\mbox{\,\footnotesize syn},L}_{\ell}$ 
and $C^{\mbox{\,\footnotesize syn},uL}_{\ell}$, respectively.

\begin{table*}
 \caption {The cosmological parameters used to generate the synthetic angular power spectra.} \label{table1} 
   \begin{center}
    \begin{tabular}{|l|l|l|l|l|l|l|}
        \hline 
         parameters  & $\Omega_b h^2$ & $\Omega_c h^2$ & $100 \theta_{MC}$ & $n_s$ & $\ln(10^{10}A_s)$ & $\tau$\\
         \hline \hline
         $\Lambda$CDM   & 0.022383 & 0.12011 & 1.040909 & 0.96605 & 3.0448 & 0.0543 \\
         \hline
  \end {tabular}
 \end{center}
\end {table*}

\item We then vary the value of the lensing amplitude $A_{lens}$ and construct the quantity $\delta^{\mbox{\,\footnotesize exc}}_{\ell}(A_{lens})$. 
\end{itemize}
 Moreover the relative error for $\delta^{obs}_{\ell}$,  $error(\delta)_{\ell}$ ,  is calculated by considering the error 
 from the observed Planck TT APS, 
$\sigma^{\mbox{\footnotesize Planck}}_{\ell}$, 
and using the standard approach for propagation of uncertainties  from equation (\ref{Dobs}).

Our analysis shall consider binnation inspections beside that one adopted by the Planck team. 
Thus, after calculating the spectrum difference 
for each $A_{lens}$ considered, and the corresponding errors,  we have to choose 
the bin length $\Delta \ell$, which is the size of the bin that we are dividing our $\delta^{obs}_{\ell}$ array into. 
Each bin will be represented  by its  mean value  with error $\sigma_{\ell} \equiv error(\delta)_{\ell} / \sqrt{\,\Delta \ell}$.
In the last step, we  compute the $\chi^2$, after binning the spectrum difference and computing the error 
\begin{equation}\label{chi}
\chi^2(A_ {lens}) =  \sum\frac{[\delta^{\mbox{\,\footnotesize obs}}_{\ell} - \delta^{\mbox{\,\footnotesize exc}}_{\ell}(A_ {lens})]^2}{\sigma^2_{\ell}} \,,
\end{equation}
where $\delta^{obs}_{\ell}$ and $\delta^{exc}_{\ell}(A_{lens})$ are defined in equations (\ref{Dobs}) and (\ref{Dsyn}), respectively. 
The sum above is performed over the binned spectra difference data $\{ \delta_{\ell}^{\mbox{\,\footnotesize obs / exc}} \}$.






Finally,  the space of parameters is extended to include the
neutrino mass $\sum m_\nu$ and the spatial curvature $\Omega_k$  to explore the possibility that any lensing excess signal can be  mimicked by  the effects of some of these parameters.

\newpage
\subsection{Null hypothesis analyses}\label{sec3.2}

In this subsection we shall test if the set of values $\{ \delta^{obs}_{\ell} \}$ is a statistical noise or not, 
i.e. we test the randomness of the spectrum difference. 
To perform this test we consider the null hypothesis: 
\\
\\
\noindent
$H_0$: The spectrum difference, $\{ \delta^{obs}_{\ell} \}$, corresponds to a residual statistical (or white) noise. 
\\
\\
That $H_0$ be true means that $A_L= 1$ completely accounts for with the lensing signal in the observed Planck APS. 
In order to examine this hypothesis we shall use the Ljung-Box (LB) test~\citep{ljungbox}, which is a modification 
of the Box-Pierce Portmanteau {\bf Q} statistic~\citep[][]{boxpierce}. 
The LB test is used to look for correlation in a data series,  
determining whether there is or not  a remaining signature in the residuals after a forecast model has been fitted to the data. 
Basically, the LB test is a useful tool to evaluate the autocorrelation between the data in analysis, and to quantify its 
statistical significance.

As a first step, it is necessary to compute the autocorrelation in a given data set 
$\{ \delta^{obs}_{i} \}$ in an interval  $M = [\ell_{min},\ell_{max}]$ with $N$ data points
\begin{eqnarray} \label{auto}
  \rho_k =\frac{\sum^{N-k}_{i=1} \left( \delta^{obs}_{i}- \overline{\delta^{obs}_{i}}\right)\left(\delta^{obs}_{i+k}- \overline{\delta^{obs}_{i}}\right)}{\sum^N_{i=1} \left(\delta^{obs}_{i}- \overline{\delta^{obs}_{i}}\right)^2} \,,
\end{eqnarray}
where $\overline{\delta^{obs}_{i}}$ is the average of all $N$ points in the $M$ interval, $k$ is commonly called the  \text{lag} and $\rho_k$ is called the lag $k$  autocorrelation.

Since $\rho_k$ measures the correlation between multipoles separated by $k$, the autocorrelation $\rho_k$ can be used, in principle, to detect non-randomness in the data.  However, it is more recommended to use tests considering multiple (sometimes called global or total) correlations across all the data interval and for  several lags jointly, like the LB test~ \citep{ljungbox}. 
The null hypothesis $H_0$ for this test establishes that the first $h$ lags autocorrelations are jointly zero, i.e.
\begin{eqnarray}
H_0: \rho_1 = \rho_2 = \cdots =\rho_k= \cdots =\rho_h = 0 \,,
\end{eqnarray}
where $h$ is the maximum lag considered in the test. In other words, $H_0$ being true implies that all the analyzed data are uncorrelated and correspond to a white noise signal. The LB statistic is defined by
\begin{eqnarray} \label{q}
   Q_h = N(N+2)\sum^h_{k=1} \frac{\rho^2_k}{N-k} \,,
\end{eqnarray}
where $\rho_k$ is the estimated correlation using equation (\ref{auto}). 
Thus , LB test does not consider just a particular lag $k$ but a set of $h$ estimated correlations.

Since $Q$ asymptotically follows a $\chi^2$ distribution,  to determine the statistical significance of the test 
it is compared to a $\chi^2$ distribution with $h'=h-q$ degrees  of freedom under the condition \citep{ljungbox} 
\begin{eqnarray}
Q_h > \chi^2_{1-\alpha,h'} \,,
\end{eqnarray}
where $q$ is the number of parameters used to fit the observed Planck APS and $\alpha$ is the significance level. 
Then, small $p$-values  ($p<\alpha$) will imply that  
significant correlation exists between the data in the set $\{ \delta^{obs}_{i} \}$  and thus, $H_0$ is rejected in the $M$ interval. 

\newpage
A choice of the $h$ parameter in equation (\ref{q}) requires a more detailed discussion. 
Several studies has been performed to define which is the optimal value. 
For instance, empirically \cite{ljung1986} suggests $h=5$, \cite{Tsay2010} suggests $h\sim \ln N$, \cite{Hyndman2018} 
$h = min(10,N/5)$, \cite{Shumway2011} $h=20$. Furthermore, \cite{Hyndman2014}  employed a study with simulations to show that
for very large values of $h$, LB test could lead to  not so reliable results. 
Recently, \cite{hassani2020}  also
used simulations to evaluate the optimal value for the number of lags $h$ involved in the LB test. 
Their results have shown that for the order of thousands data, the optimal values are $h=50$ for $\alpha=0.05$ and $h=25$ for $\alpha = 0.01$. 

To perform the LB test we first compute $\{ \delta^{obs}_{l} \}$ in different $M$ intervals and apply it for each interval. 
Our results are summarized in table~\ref{LBIntervals} where $q=7$, $\alpha =0.01$ and, as recommended by~\cite{hassani2020}, $h =25$ was used. 
We rerun the test for different values of $h$ mentioned in the previous paragraph and confirm  that the results in the table~\ref{LBIntervals} are robust. 
The first interval analyzed is $2\leq \ell \leq 2500$, i.e. the complete range of multipoles measured by Planck. 
In such interval we found a rejection of null hypothesis $H_0$ with $p = 0.0$. Then, we applied the test for three different intervals: 
$2 \leq \ell \leq 100$, $2\leq \ell \leq 800$ and $2 \leq \ell \leq 1200$, where it was found a no rejection of $H_0$. 
These results are indicating that the CMB power spectrum is well fitted by the cosmological parameters found by Planck team, and thus residual $\{ \delta^{obs}_{\ell} \}$ is white noise until, roughly, $\ell \sim 1200$. 
We also considered three more intervals, 
$1100\leq \ell \leq 2500$, $1600\leq \ell \leq 2500$ and $2000 \leq \ell \leq 2500$. 
Results in such intervals are indicating a rejection of the $H_0$ hypothesis at more than $99\%$ CL, i.e. 
$\{ \delta^{obs}_{\ell} \}$ is not white noise and some signal could be hidden in such multipole intervals. 
One possibility is that rejection of $H_0$ would be due to the not so accurate measurements, represented by the large error bars, observed in the last multipoles (especially for 
$\ell \gtrsim 2200$). 
In order to evaluate this possibility we consider one more interval.


\begin{table}
\caption {Statistical LB analyses to investigate the rejection or acceptance of the null hypothesis $H_0$}
\begin{center}
\begin{tabular}{|l|l|l|l|}
\hline 
Multipole intervals & $p$-value & Reject $H_0$ & $\%$repetition \\
\hline \hline
  $\ell = [2, 2500]$ & $0$ & Yes & $100 \%$  \\
  \hline
  $\ell = [2, 100]$ & $6.0346 \times 10^{-1}$ & No & $95 \%$  \\
\hline
  $\ell = [2, 800]$ & $1.1739\times 10^{-1}$ & No & $56 \%$ \\
\hline
  $\ell = [2, 1200]$ & $1.7097\times 10^{-2}$ &  No & $20 \%$  \\
\hline
  $\ell = [1100, 2500]$ & $1.0436\times 10^{-14}$ & Yes & $100 \%$ \\
\hline
  $\ell = [1600, 2500]$& $3.7059\times 10^{-8} $ &  Yes & $94 \%$  \\
\hline
  $\ell = [2000, 2500]$ & $1.5677\times 10^{-3}$ &  Yes & $59 \%$  \\
  \hline
  $\ell = [1100, 2200]$ & $2.6334\times 10^{-13}$ & Yes & $100 \%$ \\
  \hline
\end{tabular} 
\end{center}\label{LBIntervals}
\end{table}

In fact, we also consider the statistical LB analyses in the interval: $1100 \leq \ell \leq 2200$, and according to the very small $p$-values obtained,  we also confirm the rejection of the null hypothesis $H_0$ in this interval at $> 99\%$ confidence level (CL) as we can see in the table \ref{LBIntervals}. 
These results confirm a significant correlation among the $\{ \delta^{obs}_{\ell} \}$ and support the hypothesis of the 
presence of some structure left in the spectrum difference, even if one does not consider the data with largest errors 
for $\ell > 2200$. 
Roughly, the null hypothesis is rejected for $\ell \gtrsim 1000 -1200$, precisely at the angular scales where theoretical estimates 
indicate that lensing signal starts to be more relevant \citep{Lewis-Challinor}. 

So far, we have applied the LB test to the main values of the spectrum difference, i.e. without considering the errors $error(\delta)_{\ell}$. 
\newpage
Now, let us investigate the robustness of our results by introducing  error measurements in the LB test. In order to continue the analyses we construct artificial sets $\{{\widetilde\delta}^{obs}_{\ell}\}$ defined as  $\{ {\widetilde\delta}^{obs}_{\ell}\} \equiv \{ \delta^{obs}_{\ell} + R_{\ell} \}$ and generated as follows: first, for each $\ell$ we generate a $R_{\ell}$ which is a random number obtained from a Gaussian distribution --with zero mean and standard deviation: $error(\delta)_{\ell}$-- that is added to $\{ \delta^{obs}_{\ell}\}$ to get $\{{\widetilde\delta}^{obs}_{\ell}\}$. 
These sets $\{{\widetilde\delta}^{obs}_{\ell}\}$ are hypothetical points representing simulated spectrum differences inside 1$\sigma$ error. 
We then apply the LB test, in different M intervals, for $100$ simulated spectrum differences $\{{\widetilde\delta}^{obs}_{\ell}\}$ and compute the percentage of repetition of rejection, or not rejection, of the null hypothesis. 
The results are  shown in the last column in table \ref{LBIntervals}.

For example, in the first interval $2 \leq \ell \leq 2500$, the $100\%$ of repetitions reject $H_0$, while in the second interval the  
$95\%$ of repetitions do not reject $H_0$. The percentage of repetitions for $H_0$ no rejection is decreasing for the next two intervals: $2 \leq \ell \leq 800$ and $2 \leq \ell \leq 1200$, which can be an indicative that around $\ell \sim 1000$  there exists the probability of some correlation between the data $\{\delta^{obs}_{\ell}\}$. Nevertheless in all the other cases, percentage of repetitions for rejection, or no rejection, of $H_0$ is sufficiently high to reinforce the results shown in the third column of table~\ref{LBIntervals}.

As a conclusion of this subsection we must point out that there exist a strong evidence to reject the null hypothesis  in the interval $1100 \leq \ell \leq 2500$, where lensing effects on CMB photons are more relevant~\citep{Lewis-Challinor}. 
Nevertheless, to be conservative, we can leave out the last multipoles due to their large error bars, i.e., $\ell > 2200$, 
and will focus our next analyses in the interval $1100 \leq \ell \leq 2200$. 

Clearly, the CMB temperature data have noise, that is not isotropic due to the observational strategy of the Planck probe. 
However, the conclusion of this subsection remains roubust because in Appendix~\ref{noisemaps} we show the results from complementary 
analyses considering the APS of Planck noise maps (from two frequencies: 545 and 857 GHz). 
Moreover, in the current LB tests we consider the best-fit values from table~(\ref{table1}), but not their corresponding errors. 
Analyses taking into account the error bars, done in Appendix~\ref{varcosmopar}, 
show that our original LB test results remain the same.

\subsection{Measuring the excess of lensing power}\label{sec3.3}

\begin{figure}
\centering
\includegraphics[height=3cm, width=8.5cm]{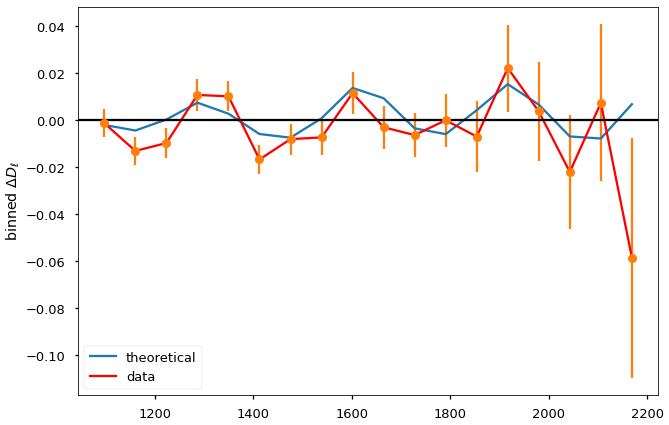}
\includegraphics[height=3cm, width=8.5cm]{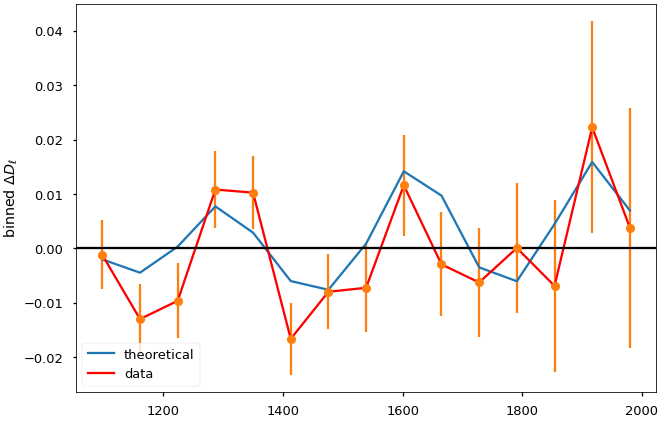}
\includegraphics[height=3cm, width=8.5cm]{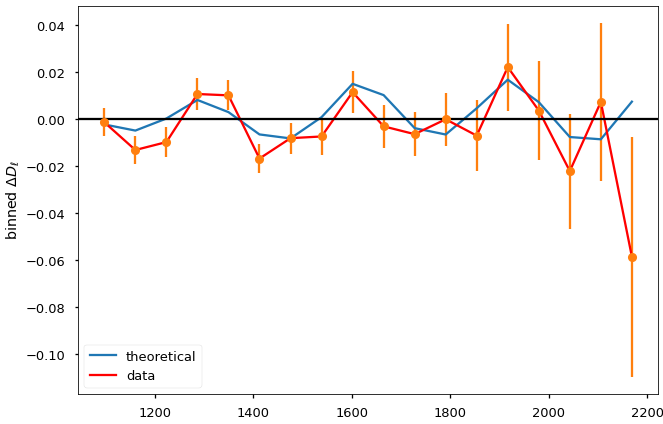}
\includegraphics[height=3cm, width=8.5cm]{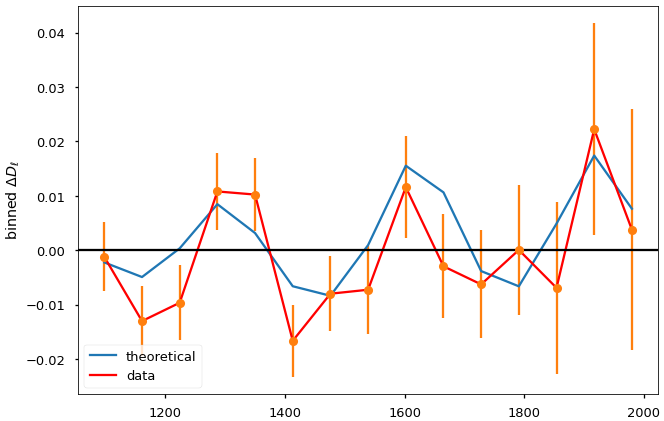}
\caption{Illustrative examples: plots of the spectrum difference $\delta^{obs}_{\ell}$ together 
with $\delta^{exc}_{\ell}(A_{lens})$, for the cases $A_{lens} = 0.20$ (first and second panels) and $A_{lens} = 0.22$ 
(third and fourth panels), for $\ell \in [1100,2000]$ and $\ell \in [1100,2200]$ as indicated in the plots.
In all these plots we consider $\Delta \ell = 63$. 
Although the best-fit curve in the first and third panels seem equals, 
they are indeed slightly different.
}\label{fig1}
\end{figure}

\begin{figure}
\centering
\includegraphics[height=3cm, width=8.5cm]{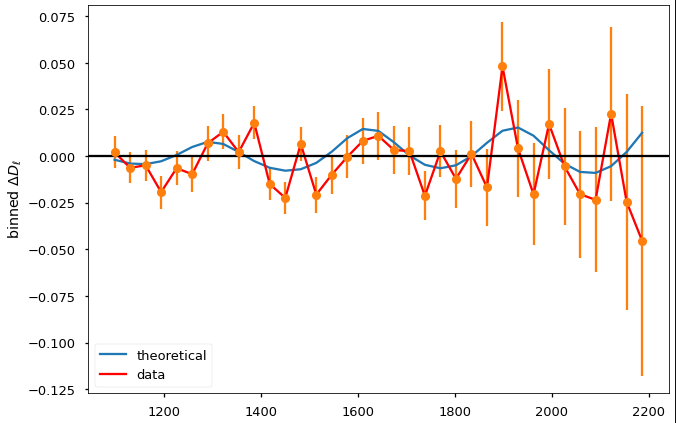}
\includegraphics[height=3cm, width=8.5cm]{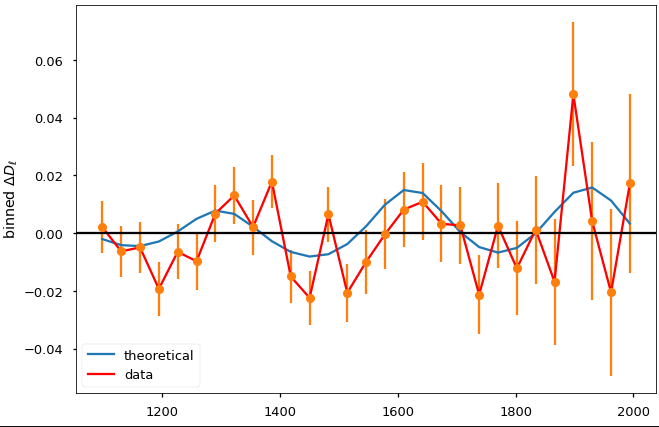}
\includegraphics[height=3cm, width=8.5cm]{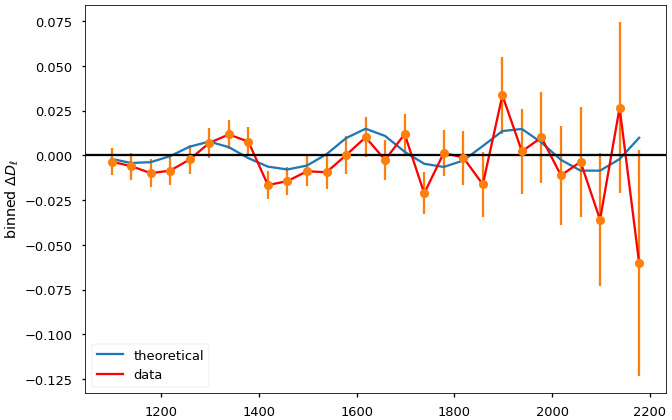}
\includegraphics[height=3cm, width=8.5cm]{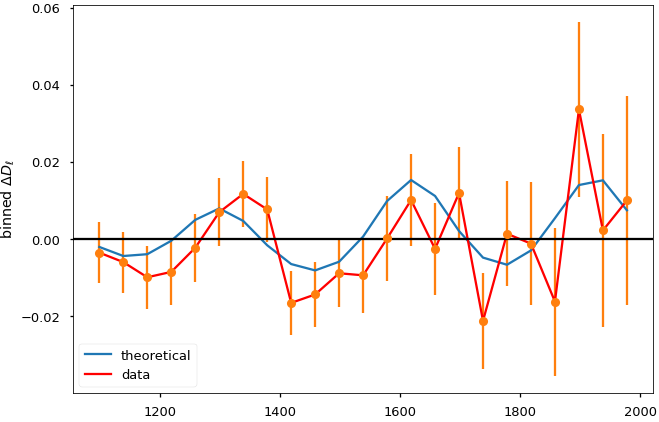}
\caption{Illustrative examples: plots of the spectrum difference $\delta^{obs}_{\ell}$ together with 
$\delta^{exc}_{\ell}(A_{lens})$ for $\ell \in [1100,2000]$ and $\ell \in [1100,2200]$ as indicated in the plots. 
In all these plots we consider $A_{lens} = 0.20$. 
They were obtained for the bin lengths $\Delta \ell = 32$, for the first and the second panels and 
$\Delta \ell = 40$ for the third and the fourth panels.
}
\label{fig2}
\end{figure}

\begin{figure}
\centering
\includegraphics[height=6cm, width=8.5cm]{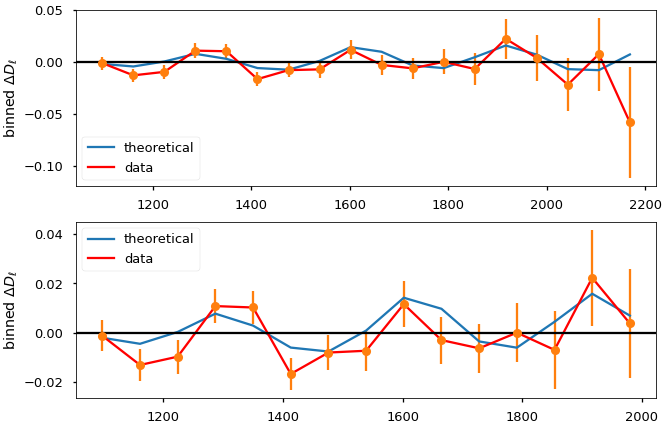}
\includegraphics[height=6cm, width=8.5cm]{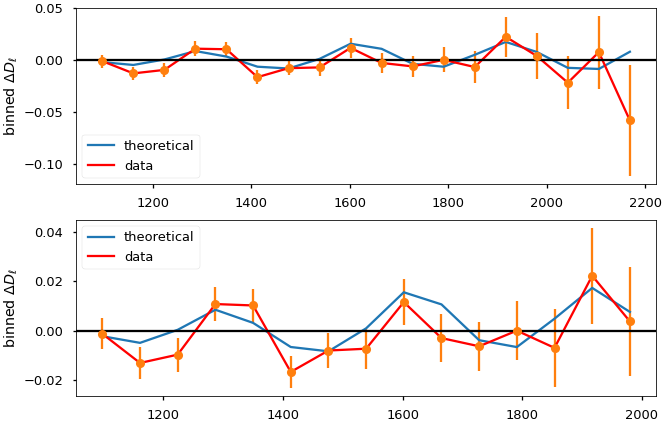}
\caption{Illustrative analyses comparing the spectrum difference $\delta^{obs}_{\ell}$ and 
$\delta^{exc}_{\ell}(A_{lens})$ for $\ell = [1100, 2200]$ and $\ell = [1100, 2000]$ with $\Delta \ell = 63$ 
for $A_{lens} = 0.20$ (upper 2 plots) and $A_{lens} = 0.22$ (lower 2 plots)}
\label{fig3}
\end{figure}

The first aim  of this work was to investigate if the spectra difference $\delta^{obs}_{\ell}$ defined in 
equation~(\ref{Dobs}), corresponds to statistical noise  or not. 
Results of the previous subsection showed that for several intervals with $\ell \gtrsim 1000-1200$ 
there seems to be a rejection of the null hypothesis $H_0$. 
Now, our second aim is to explore the possibility that the $\delta^{obs}_{\ell}$ data could be well 
reproduced through the $\chi^2$ best-fit analyses, described in the subsection~\ref{sec3.1}, 
by a synthetic APS according to the equation~(\ref{Dsyn}), 
$\delta^{\mbox{\,\footnotesize exc}}_{\ell}(A_{lens})$, for some value $A_{lens} > 0$. 
As illustrative examples for several $A_{lens}$ cases and diverse binnation schemes $\Delta \ell$, 
we show some plots of these best-fit analyses in figures~\ref{fig1}, 
\ref{fig2}, and~\ref{fig3}. 
The fact that one can find a good fit, i.e. $\chi^2 \simeq 1$ as seen in table~\ref{table3}, for the spectrum difference $\delta^{obs}_{\ell}$ 
for a synthetic APS with $A_{lens} > 0$ suggests that the lensing amplitude parameter in the Planck APS should be 
indeed larger than 1, i.e. $1 + A_{lens} > 1$, meaning that this parameter was underestimated 
in the analyses of the APS, $C^{\mbox{\footnotesize Planck}}_{\ell}$, done by the Planck collaboration. 

Assuming that the spectrum difference $\delta^{obs}_{\ell}$ has the signature of the lensing phenomenon, 
as suggested by the plots shown in figures~\ref{fig1}--\ref{fig3}, 
one can find the parameter value $A_{lens}$ of the synthetic APS that best-fits these data. 
In the table~\ref{table3} we display the $\chi^2$ values obtained for different  $A_{lens}$ 
values and diverse $\Delta \ell$ bin lengths. 
One notice that in several cases, $\chi^2 \simeq 1$ is fully possible, including the $A_{lens} = 0$ case which 
is also considered in our analyses. 
This suggests that a more detailed likelihood analyses are in due.

\begin{figure}
\centering
\includegraphics[height=6cm,width=8.5cm]{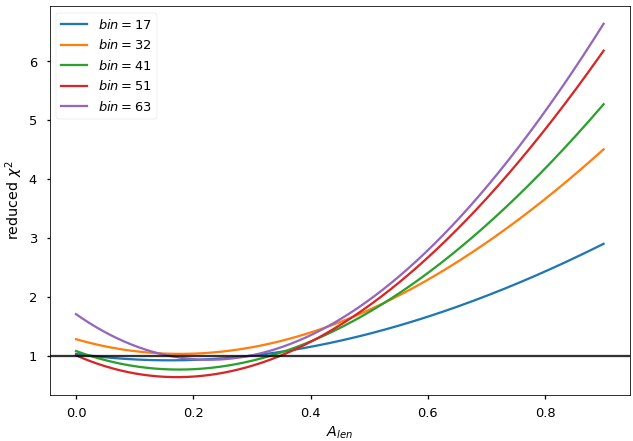}
\caption{$\chi ^2$ as a function of $A_{lens}$ and for different number of bins. This plot shows 
that, independent of the bin size $\Delta_{\ell}$, the $\chi^2$ function exhibits solutions for the case $\chi^2 = 1$ for 
$A_{lens} \ne 0$, which justify our search for the value $A_{lens}$ that best-fits the observed data 
$\delta^{obs}_{\ell}$. 
}
\label{fig4}
\end{figure}

\begin{figure}
\centering
\includegraphics[height=7cm, width=8.5cm]{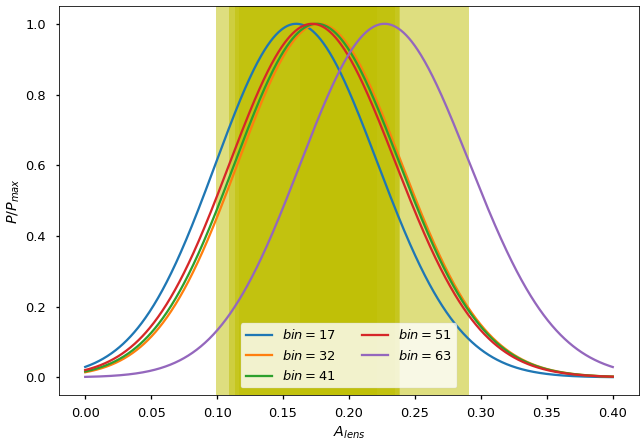}
\caption{Maximum likelihood analyses, complementing the analyses done in figure~\protect\ref{fig4}, showing the 68\% CL of the $A_{lens}$ values that perform the reduced $\chi^2$ best-fit of the spectrum difference $\delta^{obs}_{\ell}$, that is $A_{lens} \in [0.10,0.29]$, for several bin sizes.}
\label{fig5}
\end{figure}

\begin{figure}
\centering
\includegraphics[height=7cm, width=8.5cm]{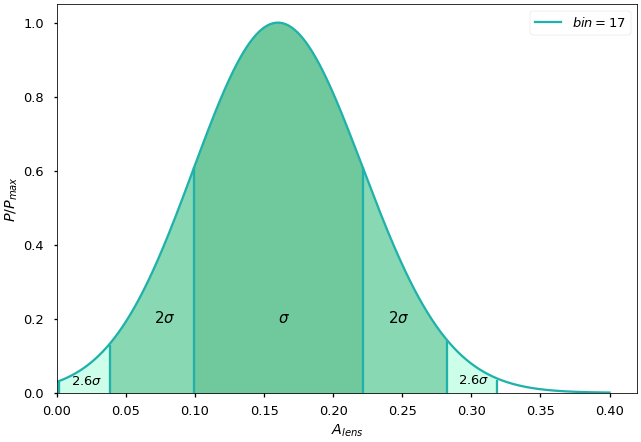}
\caption{Illustrative example of $\sigma$ intervals for the $\Delta \ell = 17$ case. 
As observed, for this binnation case the value $A_{lens} = 0$ (equivalently  $A_{L} = 1$) 
is excluded with $2.6 \sigma$ (i.e., at 99\% CL).}
\label{fig6}
\end{figure}

We start considering the $\chi^2$ procedure, of the synthetic APS that best-fits the spectrum difference, $\delta^{obs}_{\ell}$, 
for a continuous set of $A_{lens}$ values. 
As a result, one obtains a $\chi^2$ value for each $A_{lens}$ considered, that is, $\chi^2$ becomes a function of the $A_{lens}$ 
parameter: $\chi^2 = \chi^2(A_{lens})$, information that can be plotted as curves $\chi^2$ versus $A_{lens}$, as observed 
in the figure~\ref{fig4}, considering moreover different bin lengths $\Delta \ell$. 
In the table~\ref{table4} we show the numerical values of the intersections between the curves $\chi^2(A_{lens})$ 
appearing in figure~\ref{fig4}, obtained for different binnations, and the straight line 
$\chi^2 = 1$. 
According to these results, we conclude that values $A_{lens} \ne 0$ exist, which imply that the spectrum difference 
$\delta^{obs}_{\ell}$ can be  well fitted by  $\delta^{exc}_{\ell}(A_{lens})$ for some $A_{lens} \ne 0$ values and several $\Delta \ell$ choices. 
Furthermore, we perform statistical likelihood analyses to find the best-fit $A_{lens}$ values, as shown in table~\ref{table5}, also  for several bin lengths. 
Observing the figure~\ref{fig4} one can argue that several binnations provide a best-fit with reduced $\chi^2 = 1$ 
for some $A_{lens} = A_{lens}(\Delta \ell)$ values, meaning that a solution with $A_{lens} \ne 0$ exists but is not unique. 
Complementing this $\chi^2$ studies, the maximum likelihood analyses shown in figure~\ref{fig5} let us to conclude 
that $A_{lens} \in [0.10,0.29]$ at 68\%~CL. See also the figure~\ref{fig6} for a detailed study of the illustrative case $\Delta \ell = 17$, where it can be seen that $A_{lens} > 0$ at 2.6 $\!\sigma$. 
The same results are found for other binnation schemes. 

As a result of the analyses in this subsection we found that the signal hidden in $\{\delta^{obs}_{\ell}\}$ is related to an 
excess of lensing amplitude. 
However, we reached this conclusion by considering only the best-fit values of table~(\ref{table1}). 
It is then necessary to test for a possible bias with this choice, and for this we shall perform  complementary analyses considering the measured error bars of the best-fit cosmological parameters. 
Results of these tests are discussed in the Appendix~\ref{varcosmopar}, where we have proven that not only the hidden signal in $\{\delta^{obs}_{\ell}\}$ is very weakly correlated with the cosmological parameters from table~(\ref{table1}), but also the fact that a $\sim 20\%$ lensing amplitude excess in $\{\delta^{obs}_{\ell}\}$ is robust.
%

\newpage



\begin{table}
\caption{Illustrative examples of $\chi^2$ calculations for different values of $A_{lens}$ 
and several binnation choices $\Delta \ell$. Some of these cases can be seen in figures~\ref{fig1},~\ref{fig2}, 
and~\ref{fig3}.} \label{table3} 
\begin{center}
\begin{tabular}{|l|l|l|l|l|l|l|l|}
\hline 
$A_{lens}$ & $\Delta \ell = 17$ & $\Delta \ell = 32$ & $\Delta \ell = 41$ & $\Delta \ell = 51$ & $\Delta \ell = 63$ \\
\hline \hline
0.00 &1.0363 & 1.2809 & 1.02590 & 0.95802 & 1.63446 \\
\hline
0.10 &0.9414 &  1.0747 & 0.7610 & 0.6416 & 1.0798 \\
\hline
0.20 & 0.9329 & 1.0305 & 0.7094 & 0.5851 & 0.8395 \\
\hline  
0.24 & 0.9525 & 1.0561 & 0.7459 & 0.6319 & 0.8274 \\
\hline  
0.34 & 1.0565 & 1.2238 & 0.9734 & 0.9150 & 0.9980 \\
\hline
\end{tabular} 
\end{center}
\end{table}

\vspace{0.5cm}

\subsection{Cosmological parameters dependency}

In the previous section we explore the possibility that the residual structure in CMB lensing signal can be explained by an excess in lensing  amplitude $A_{lens} > 0$. However, there is also the possibility that such residual structure could be mimicked by varying in other cosmological parameters. The scrutiny of this dependency is important to confirm if the signal found in the previous section is indeed due to 
the lensing effect or it also has contribution of other cosmological parameters. 
This analysis is the main objective in this subsection.

As it is well known, the APS, $C^{\Lambda\mbox{\footnotesize CDM}}_{\ell}$, is sensitive to the cosmological parameters \citep{Peebles1968,Doroshkevich1978,Wilson1981}. Such sensitivity is inherited by the quantity $\delta^{obs}_{\ell}$ defined in equation  (\ref{Dobs}). 
Despite that, although the 
cosmological parameters in table~\ref{table1} have a high impact in the CMB APS, this is not the case when the spectrum difference is considered, where their impact is weaker. In some sense, the analysis of $\delta^{obs}_{\ell}$ will diminish possible  degeneracies between those cosmological parameters and $A_{lens}$, $\sum m_\nu$, and $\Omega_k$.

So far, the model considered was the simplest $\Lambda$CDM  with the six parameters listed in table~\ref{table1} plus lensing amplitude $A_L=1$. Now, the space of parameters is extended to include the neutrino mass $\sum m_\nu$ and the spatial curvature $\Omega_k$. Massive neutrinos slow down the growth of matter perturbations and prevent clustering process \citep{Bond1980,Lesgourgues2006}, resulting, therefore, in an anticorrelation with gravitational lensing amplitude. Moreover, spatial curvature also has correlation with lensing and thus, CMB lensing signal  is sensitive to the neutrino mass and the spatial curvature of the universe. 


In order to investigate the space of parameters 
($A_{lens}$, $\sum m_\nu$, $\Omega_k$) in the light of the spectrum difference  $\delta^{obs}_{\ell}$, the other cosmological parameters were fixed by using  best-fit values  found by Planck 2018 data~\cite{Planck18} (see table \ref{table1}) and the parameters $A_{lens}$, $\sum m_\nu$, and $\Omega_k$ were allowed to vary. Then we use MCMC techniques to find constraints on these three free parameters  by considering flat priors and marginalizing to compute the one dimensional posterior distributions. Results are presented in 
figure~\ref{fig7} and table~\ref{table7}.

In figure~\ref{fig7} the one dimensional posterior distributions for $\Delta \ell = 17$ and $\Delta \ell =63$ are shown. We can say, in general, that our results are consistent at $68 \%$ CL with an almost flat spatial curvature, with $\sum m_\nu < 0.47$ eV, and with a $\sim 20\%$ lensing amplitude excess quantified by the  $A_{lens}$ parameter. This $A_{lens}$ excess is still present at around three  standard deviations (see table \ref{table7}), even though these analyses considered a  larger number of parameters. 
All these results are robust when changing $\Delta  \ell$ indicating that  the neutrino mass or the spatial curvature weakly impact in the $\delta^{obs}_{\ell}$ intensity signal. 
Therefore their effects are not sufficient to explain the signature in the spectrum 
difference $\delta^{obs}_{\ell}$, which is more likely explained by a lensing amplitude $A_{lens} \simeq 0.20$, which implies that the lensing amplitude in the Planck's CMB APS should be $\sim 20\%$ larger than the value expected in the  flat-$\Lambda$CDM model. 


\begin{table}
\caption{Table with the numerical values obtained due to the intersection between the horizontal line, 
representing  the reduced-$\chi^2 = 1$, with the curves produced calculating the reduced-$\chi^2$ for the parameter $A_{L}$, 
and considering various bin-length cases $\Delta \ell$. 
The first and second points in the table below represent the values of $A_{L}$ where $\chi ^2 = 1$. 
As observed in figure~\ref{fig4} there is no intersection for the case $\Delta \ell = 32$. 
} \label{tab:title} 
\begin{center}
\begin{tabular}{|l|l|l|l|l|l|l|l|}
\hline 
bin length & first point & second point \\
\hline \hline
$\Delta \ell = 17$ & $0.028675$ & $0.29479$ \\
\hline
$\Delta \ell =32$ & \hspace{0.4cm}-- & \hspace{0.4cm}--  \\
\hline
$\Delta \ell =41$ & $0.02495$ & $0.3305$ \\
\hline  
$\Delta \ell =51$ & $0.0023$ & $0.3477$ \\
\hline  
$\Delta \ell =63$ & $0.16053$ & $0.29425$ \\
\hline
\end{tabular} 
\end{center}\label{table4}
\end{table}

\begin{table}
\caption{Values of $A_{L}$ from the likelihood plot. The last column indicates the confidence level (CL) for having $A_{lens} > 0$ in each binnation case.}~\label{table5}
\begin{center}
\begin{tabular}{|l|c|c|c|l|l|l|l|}
\hline 
$\!$bin length$\!$ & $\!\!A_{lens}$ at max. likelihood$\!$ & $\!\!$1$\sigma$ interval$\!$ & $\!\!$CL for $A_{lens}\!>\!0$$\!\!$ \\
\hline \hline
$\Delta \ell = 17$ & $0.16036$ & $[0.0993, 0.2213]$ & 2.6$\sigma$\\
\hline
$\Delta \ell =32$ & $0.177477$ & $[0.1164, 0.2384]$ & 2.8$\sigma$\\
\hline
$\Delta \ell =41$ & $0.17567$ & $[0.1136, 0.2376]$  & 2.8$\sigma$\\
\hline  
$\Delta \ell =51$ & $0.172072$ & $[0.1090, 0.2350]$ & 2.7$\sigma$\\
\hline  
$\Delta \ell =63$ & $0.22702$ & $[0.1630, 0.2910]$ & 3.5$\sigma$\\
\hline
\end{tabular} 
\end{center}\label{likelihoods}
\end{table}

\section{Discussions and Final Remarks}\label{sec4}

We now discuss the exhaustive analyses done  with the precise Planck APS data and the results obtained investigating the statistical 
features of the spectrum difference $\delta^{obs}_{\ell}$. 
Recent analyses have reported the preference of the lensing amplitude parameter $A_L$ for a value larger than 1, with high statistical significance~\citep{Planck16,Planck18,Bianchini,DES22}, 
and we want to study if this result could be sourced by a  statistical artifact or if it has a physical origin. 
We emphasize that our analyses must be intended as complementary (due to the methodology employed) and independent 
(with respect to previous works and those done by the Planck collaboration~\cite{Planck18}). 
%

In this respect, it is interesting that the underestimated lensing amplitude has also been observed in the lensing-potential power spectrum, 
$C_{\ell}^{\phi \phi}$, as shown in~\cite{Planck18}, where it was measured $A_L = 1.19$ for the best-fit model 
(see the discussion in section 2.3  in ref.~\cite{Planck18}), 
a value in perfect agreement with the result of our analyses.

On the other hand, we discuss some interesting issues regarding the study of the lensing amplitudeproblem that are not usually discussed 
in the literature. 
Firstly, in the usual statistical
analyses of the CMB APS, the signature left in the spectrum difference is not discussed, for this we consider that previous to any best-fitting model procedure considering the lensing amplitude as a parameter, one needs to discuss the nature of the spectrum difference leftover: : is it a residual white noise, or not? Second, because we find that the spectrum difference has a definite signature (i.e., it is not white noise), the binnation procedure may have an impact in the measurement of$A_{lens}$, and therefore deserves to be analyzed . Third, since the spectrum
difference has a definite signature, not any parameter –or combination of parameters– can
suitably best-fit it, a fact that when is considered contributes to eliminate, or at least diminish,
possible degeneracies (because each cosmological parameter has a different effect in the fitting
of$\delta^{obs}_{\ell}$).

In the previous section we have performed a detailed examination of the unbinned CMB APS. 
We have proved that the spectrum difference $\delta^{obs}_{\ell}$ is not statistical noise; 
additionally we found several multipole intervals, starting at $\ell \gtrsim 1000$, where this result is true 
(see table~\ref{LBIntervals}), supporting the crucial  point that the spectrum difference is not statistical noise at small scales. 
This result is suggesting that some signal is present in the spectra difference, and the signature shown in 
these data must be an indication of the phenomenon that caused it. 
However, degeneracy can also happen, that is, the signature present in the data could be reproduced by more than 
one source and for this reason we also explore diverse hypotheses to explain the signature and the amplitude of 
the spectrum difference data.

Bearing in mind that, at the scales $\ell \gtrsim 1000$, the acoustic peaks are extremely sensitive to the lensing 
phenomenon, the first hypothesis examined was that the signature in the spectrum difference corresponds to an underestimated lensing amplitude in the Planck best-fitting procedure. 
As shown in the analyses of section~\ref{sec3.3} this hypothesis is indeed verified, where we find a lack of lensing amplitude of around $20\%$ with respect to the Planck APS fitted assuming the flat-$\Lambda$CDM model with $A_L = 1$. 
As a matter of fact, we find that there is an excess of signal in $\delta^{obs}_{\ell}$ (see equation~(\ref{Dobs})) 
that is well explained by a $\Lambda$CDM APS with a non-null lensing amplitude parameter $A_{lens} > 0$, with values in the interval 
$[0.10,0.29]$ at 68\%~CL. 
Moreover, we found several scheme binnations that best-fits the spectrum difference $\delta^{obs}_{\ell}$, 
with $\chi^2 = 1$, with lensing amplitudes $A_{lens} \simeq 0.2$ (see 
tables~\ref{table4} and~\ref{table5}). 
According to our likelihood analyses, the synthetic APS produced for these best-fit procedures of $\delta^{obs}_{\ell}$, 
with lensing amplitude $A_{lens}$ as a parameter, show that $A_{lens} > 0$, or equivalently $A_L > 1$, with statistical significance of $\sim 3 \sigma$. 

\begin{figure}
\centering
\includegraphics[height=16cm, width=8.5cm]{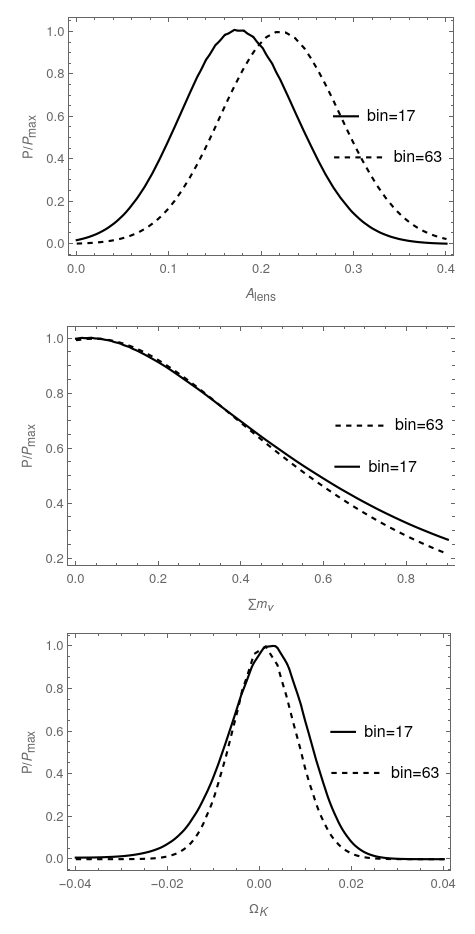}
\caption{One dimensional posterior distributions for $A_{lens}$, $\sum m_\nu$ and $\Omega_k$ for $\Delta \ell=17$ and  $\Delta \ell=63$, as illustrative examples. 
The $\delta^{obs}_{\ell}$ signal is consistent with: $\sum m_\nu < 0.47$eV, an almost flat spatial curvature, and $\sim 20\%$ of excess in $A_{lens}$. These results show the consistency of the different values that these  parameters can assume to explain the spectrum difference $\delta^{obs}_{\ell}$.}
\label{fig7}
\end{figure}

Additionally, we have also investigated a possible dependency of the spectrum difference $\delta^{obs}_{\ell}$, 
in signature and intensity, on some cosmological parameters associated to lensing amplitude, neutrino mass and spatial curvature. Despite neutrino mass and spatial curvature can have  impact in lensing signal, the impact in $\delta^{obs}_{\ell}$ residual is weak and cannot be enough to reproduce  the excess in $A_{lens}$, in signature and intensity. 
Our likelihood analyses for the examination of the cosmological parameters 
$A_{lens}$, $\sum m_{\nu}$, and $\Omega_k$ are displayed in figure~\ref{fig7}.

\begin{table} 
\caption{Values $A_{lens}$, $\sum m_\nu$ and $\Omega_k$ at 1$\sigma$ for $\Delta \ell = 17$ and  $\Delta \ell = 63$. 
The last column indicates the confidence level (CL) for having $A_{lens} > 0$ in each binnation case.}~\label{table7}
\begin{center}
\begin{tabular}{|l|c|c|c|l|l|l|l|}
\hline 
$\!$bin length$\!$ & $\!A_{lens}$ $\!$ & \!$\sum m_\nu $\!  & $\!\Omega_K\!$ & \!\!CL for $A_{lens}\!>\!0$\\
\hline \hline
$\Delta \ell = 17$ & $0.18^{+0.063}_{-0.063}$ & $<0.47$ eV & $0.0029^{+0.009}_{-0.009}$ & 3$\sigma$\\
\hline
$\Delta \ell =63$ & $0.22^{+0.065}_{-0.065}$ & $< 0.45$ eV & $0.0013^{+0.0071}_{-0.0071}$ & 2.9$\sigma$\\
\hline
\end{tabular} 
\end{center} 
\end{table}

\newpage
\appendix
\section{Appendix: Robustness Tests}

We perform a variety of robustness tests to support the validity of our results. 
The first set of tests study the hypothesis of a possible bias in our results coming from the fact 
that we used the best-fit Planck cosmological parameters, displayed in table~\ref{table1}, ignoring 
their measured uncertainty. 
Because this is a possible source of bias in the production of the synthetic CMB APS, we repeat our 
analyses constructing the synthetic maps with cosmological parameters randomly chosen 
from the $1 \sigma$ confidence interval (table 2 of \cite{Planck18}). 
The second set of tests investigate the possibility that a residual noise in the CMB temperature 
maps could have some impact in the Planck APS data possibly contributing to the excess of 
lensing amplitude found in our analyses.

\subsection{Variation of cosmological parameters}\label{varcosmopar}

In this section we present the results of the analyses that include the measured errors of the Planck cosmological parameters in the production of the synthetic APS. To include the errors measured by the Planck in our tests, we first generate 100 different points in the six-dimensional $\Lambda$CDM space of parameters. Then, each point consists of a set of the six cosmological parameters given in table~\ref{table1}, which were randomly generated following a normal distribution within 1$\sigma$ interval (using the table 2 of \cite{Planck18}). 
For each of the 100 sets of six parameters each, we construt 100 synthetic APS to replace the set of \,$C^{\Lambda CDM}_l$ in equation~(\ref{Dobs}), generating 
thus their correspondig 100 $\{\delta^{obs}_{\ell}\}$ sets, then the same analyses done in sec. \ref{sec3.1} and sec. \ref{sec3.2} are performed again. 

The first set of analyses is meant to test the possible existence of a bias in the acceptance/rejection of the null hypothesis $H_0$ 
in the relevant multipole interval $[1100,2200]$ caused by the inclusion of the 1$\sigma$ errors measured by the Planck Collaboration. 
To do this, we apply the LB test for each of 100 sets of  $\{\delta^{obs}_{\ell}\}$ constructed as described in the previous paragraph. 
The results show that, considering all the 100 times where $H_0$ was tested it is rejected. 
The distribution of the $p-$values obtained has a mean $\simeq 9.9 \times \sim 10^{-14}$ and 
a median $\simeq 9.6 \times 10^{-14}$, which means that even considering the 1$\sigma$ errors 
of the best-fit values of the table~\ref{table1}, the rejection of the null hypothesis is robust, supporting our original result that the residual  $\{\delta^{obs}_{\ell}\}$ signal in the multipole interval $[1100,2200]$ is not noise.
 
\begin{figure}[h!] 
\centering
\includegraphics[height=7cm, width=8.5cm]{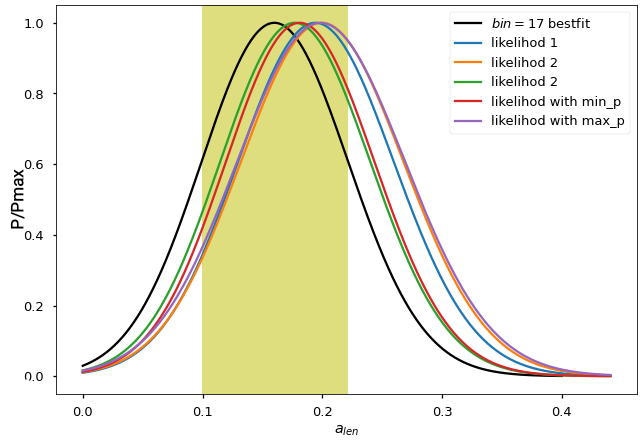}
\caption{Illustrative example of  the $\Delta \ell = 17$, likelihoods for different set of cosmological parameters.
Shadowed region represents the 1 $\sigma$ interval for the best-fit case. The maximum values of the likelihoods for the other sets 
of parameters are inside that region.} 
\label{fig8}
\end{figure}

 The second analyses in this subsection is to test whether our conclusion of a lensing excess hidden  in $\{\delta^{obs}_{\ell}\}$ signal is biased by the inclusion of the best-fit error bars. To do that, we use equation~(\ref{chi}) to compute the likelihood of the 100 $\{\delta^{obs}_{\ell}\}$ sets. All likelihoods present the same excess of lensing around $\sim 20\%$. 
 In figure~(\ref{fig8}) we present the outcoming curves from five illustratives cases: likelihood of the best-fit value; likelihoods for the maximum and minimum $p-$value from the LB test and three likelihoods randomly choosen from the set of 100 computed likelihoods. In all cases the maximum values are in agreement at 1$\sigma$. This result means that the lensing amplitude excess found in sec.~\ref{sec3.3} is robust, and the signal hidden in  $\{\delta^{obs}_{\ell}\}$ is not correlated, or at least, it is not strongly correlated with the parameters given in table~(\ref{table1}).


\subsection{Noise maps}\label{noisemaps}

In this section we investigate the possibility that a residual noise in the CMB temperature 
maps could have some impact in the Planck APS data possibly modifiying  the conclusion of 
lensing amplitude found in our analyses. To test this we consider 20 simulated Planck noise maps for each of the two 
frequencies: 857 GHz and 545 GHz to construct new simulated APS with noise, $C^{Planck+noise}_l$, which replace the \,$C^{Planck}_l$ data in 
equation~(\ref{Dobs})~\citep{Planck15-XV}~\footnote{\url{https://pla.esac.esa.int}}. 
Thus the resulting $\{\delta^{obs+noise}_{\ell}\}$ include the noise information and we apply the LB test in several multipole intervals as shown in table~\ref{LBIntervals}. 
As described in sec.~{\ref{sec3.2}}, we now simulate 100 $\{\delta^{obs+noise}_{\ell}\}$ for each noise map. Results are summarised in 
the table~\ref{LBIntervals2}, where we show the good agreement with the percentage values of rejection/acceptance previously 
found in sec. (\ref{sec3.2}) and displayed in table~\ref{LBIntervals}. The $p-$values in all analyzed intervals were very small, in particular, for the interval $[1100,2200]$, the $p-$value distribution obtained has a mean $\simeq 1.39 \times 10^{-11}$ 
and a median $\simeq 1.96 \times 10^{-13}$ for both frequencies considered here.
This confirms that the porcentage values for rejection/acceptance of the $H_0$ hypothesis originally obtained are robust 
when the CMB maps noise are considered.

\begin{table}
\caption {Statistical LB analyses to investigate the rejection or acceptance of the null hypothesis $H_0$ in the case 
that the CMB APS in study include the contribution of simulated noise from two Planck CMB frequency maps: 
$545$ GHz and $857$ GHz. 
As observed in this table, the results are the same as those obtained in table~\ref{LBIntervals}, showing that our original 
results are robust.
}
\begin{center}
\begin{tabular}{|l|l|l|l|l|l|}
\hline 
$\!\!$Multipole intervals  & Reject $H_0$ & \% for $545$ GHz & \% for $857$ GHz \\
\hline \hline
  $\ell = [2, 2500]$  & Yes  &  $100 \%$  & $100 \%$ \\
\hline
  $\ell = [2, 800]$  & No  & $65.8 \%$  &$66 \%$\\
\hline
  $\ell = [2, 1200]$&  No  & $41.4 \%$  & $42.2 \%$\\
\hline
  $\ell = [1100, 2500]$  & Yes &  $100 \%$  &$100 \%$\\
  \hline
  $\ell = [1100, 2200]$  & Yes  &  $100 \%$  &$100 \%$\\
\hline
\end{tabular} 
\end{center}\label{LBIntervals2}
\end{table}

\newpage
\section*{Acknowledgements}

We acknowledge the use of data from the Planck/ESA mission, downloaded from the Planck Legacy Archive. 
RM  and WHR thank CNPq, TWAS, and FAPES (PRONEM No 503/2020) for the fellowships and finantial support under which this work was 
carried out. 
AB acknowledges a CNPq fellowship. 
We acknowledge that our work made use of the CHE cluster, managed and
funded by the COSMO/CBPF/MCTI, with financial support from FINEP and FAPERJ, and operating
at Javier Magnin Computing Center/CBPF.

\end{document}